\documentclass[fleqn,10pt]{wlscirep}\title{Analysis of Topological Transitions in Two-dimensional Materials by Compressed Sensing}

\author[1,2]{Carlos Mera Acosta}
\author[1]{Runhai Ouyang}
\author[2,3]{Adalberto Fazzio}
\author[1]{Matthias Scheffler} 
\author[1]{Luca M. Ghiringhelli} 
\author[1,*]{Christian Carbogno}
\affil[1]{Fritz-Haber-Institut der Max-Planck-Gesellschaft, Faradayweg 4-6, 14195 Berlin-Dahlem, Germany}
\affil[2]{Institute of Physics, University of Sao Paulo, CP 66318, 05315-970, Sao Paulo, SP, Brazil}
\affil[3]{Brazilian Nanotechnology National Laboratory, CP 6192, 13083-970, Campinas, SP, Brazil}

\affil[*]{carbogno@fhi-berlin.mpg.de}

\usepackage{graphicx}
\usepackage{multicol}
\usepackage{xcolor}
\usepackage{mathtools}
    \usepackage{amsmath}    
    \usepackage{amsfonts}   
    \usepackage{amssymb}    
    \usepackage{amsbsy} 
    \usepackage{amsthm} 
    \usepackage{mathptmx}   
\usepackage{rotating}
\usepackage{bm}
\usepackage[normalem]{ulem}

\newcommand{\SM}{Supp. Mat. }

\DeclareSymbolFont{newfont}{OML}{cmm}{m}{it}
\DeclareMathSymbol{\Epsilon}{3}{newfont}{15}
\DeclareMathSymbol{\Varrho}{3}{newfont}{37}

\begin{abstract}
\underline{Q}uantum \underline{s}pin-\underline{H}all \underline{i}nsulators (QSHIs),~i.e.,~two-dimensional \underline{t}opological \underline{i}nsulators~(TIs) with a symmetry-protected 
band inversion, have attracted considerable scientific interest in recent years. In this work, we have computed 
the topological $Z_{2}$~invariant for 220 functionalized honeycomb lattices that are isoelectronic to functionalized graphene. 
Besides confirming the TI character of well-known materials such as functionalized stanene, our study identifies 45 
yet unreported QSHIs. We applied a compressed-sensing approach to identify a physically meaningful descriptor for the 
$Z_{2}$~invariant that only depends on the properties of the material's constituent atoms. This enables us to draw a ``map of
materials'', in which metals, trivial insulators, and QSHI form distinct regions. This analysis yields fundamental 
insights in the mechanisms driving topological transitions.
The transferability of the identified model is explicitly demonstrated for an additional set of honeycomb 
lattices with different functionalizations that are not part of the original set of 220 graphene-type materials 
used to identify the descriptor. In this class, we predict 74 more novel QSHIs that have not been reported in literature
yet. 
\end{abstract}

\begin{document}

\flushbottom
\maketitle
\thispagestyle{empty}

In the last decade, the experimental realization of graphene and of topological insulators has fueled interest in the fundamental physics and possible applications 
that are hosted in linear band crossings. In fact, the crossing of energy bands in condensed matter has been investigated since the very early applications of formulation of 
quantum mechanics to periodic solids~\cite{PhysRev.52.365}. Already in 1985, Volkov and Pankratov~\cite{Pankratov1985} showed that interfacing two semiconductors with mutually inverted bands can lead to massless Dirac fermions,~i.e.,~linear electronic-dispersion relations that cross~(leading to a phenomenon referred to as ``band inversion'') 
and connect conduction and valence bands. If this inversion occurs~\cite{Kane:2005gb} at time reversed pairs of reciprocal space points~$\pm\vec{k}$, the respective boundary states associated to different spins must exhibit opposite momentum, which 
in turn forbids backscattering~\cite{ReportsonProgress_79_6_066501,acs.jpclett.7b00222,annurev_2_1_31}.  \underline{Q}uantum \underline{s}pin-\underline{H}all \underline{i}nsulators (QSHI) are two-dimensional \underline{t}opological \underline{i}nsulators~(TIs)~\cite{RevModPhys.82.3045, RevModPhys.83.1057} 
that intrinsically exhibit this property~\cite{Kane:2005gb,Bernevig:2006by}. In graphene, for instance, this band inversion is driven by \underline{s}pin-\underline{o}rbit \underline{c}oupling~(SOC), which formally leads to a minute band-gap opening~\cite{Kane:2005hl,PhysRevLett.96.106802}.  In close analogy to charge pumping in the integer quantum-Hall effect~\cite{PhysRevLett.49.405}, the spin charge pumped through the edge states is quantized in QSHIs~\cite{PhysRevB.74.195312}. 
The respective integer quantum, i.e., the topological $Z_{2}$ invariant,  is 1 in QSHIs and 0 in trivial insulators. Formally, this $Z_{2}$ invariant is defined via the integral over the \underline{h}alf \underline{B}rillouin \underline{z}one (HBZ):
\begin{equation}
Z_2 = \frac{1}{2\pi}\left[\oint_{\partial\textrm{HBZ}} \mathcal{\boldsymbol{A}}(\boldsymbol{k})dl -\int_\textrm{HBZ} \mathcal{\boldsymbol{F}}(\boldsymbol{k}) d\tau \right]\mod(2)
\end{equation}
of the Berry connection~$\mathcal{\boldsymbol{A}}(\boldsymbol{k})$ and Berry curvature~$\mathcal{\boldsymbol{F}}(\boldsymbol{k})$~\cite{PhysRevLett.95.146802,PhysRevB.74.195312}. This
can be interpreted as the effective magnetic flux of a self-induced magnetic field, the Berry curvature, through the HBZ. Here, $\partial$HBZ is the contour of HBZ. 
Although theory predicted a wide variety of QSHIs~\cite{acs.jpclett.7b00222,ReportsonProgress_79_6_066501}, only few of them feature a large enough intrinsic bulk band gap at finite temperatures to allow for an experimental characterization,~e.g.,~bilayer Bi~\cite{PhysRevLett.107.136805,nphys3048} as well as HgTe/CdTe \cite{science.1133734,science.1148047,PhysRevLett.114.126802} 
and InAs/GaSb quantum wells~\cite{PhysRevLett.107.136603,PhysRevLett.100.236601}. 

So far, the computational search for new QSHIs has been a numerically costly trial-and-error process that required
to compute the $Z_2$~invariant for each individual compound~\cite{Mounet:2018gu}. Since no simple rule of thumb exists that allows to \textit{a priori} distinguish 
trivial from topological insulators~\cite{Young:2011ju}, an ``exhaustive search procedure seems out of reach at the present time''~\cite{Gresch:2017hy}. Naturally, the search for new QSHIs was thus guided by experience and intuition,~e.g.,~by focusing on 
heavy elements with high SOC~\cite{PhysRevX.1.021001,PhysRevB.84.195430,PhysRevB.89.155438,Zhou:2014de}.  For instance, 17 potential TIs could be identified by carrying out high-throughput electronic band 
structure calculations for 60,000 materials~\cite{APR.6.4.31}. In the same spirit, high-throughput studies in this
field have been performed using semi-empirical descriptors as a guidance,~e.g.,~the derivative of the band gap with no SOC with respect to the lattice 
constant~\cite{Yang:2012kx}.

In this work, we focus on functionalized 2D honeycomb-lattice materials~\cite{Lucking:2018ia,nmat4742,nl500206u}, a material class 
in which various QSHI candidates have been found already,~e.g.,~functionalized stanene~\cite{PhysRevLett.111.136804}. 
By computing the $Z_2$-invariant for 220 of these compounds from first principles we are able to identify 45 new QSHIs that have not yet been reported in literature.  
Using a recently developed statistical-learning approach~\cite{PhysRevLett.114.105503, NewJPhys.19.023017,Runhai_paper} based on compressed sensing, we then derive a two-dimensional ``map'' of these materials, in which metals, trivial insulators, and QSHIs are separated in different domains. The axes of this map are given by nonlinear analytic functions that only depend on the properties of the material's constituent atoms, but 
not on the properties of the material itself. The identified descriptor captures the character of the electronic structure, revealing that orbital interactions 
can drive a band inversion even in compounds featuring relatively light elements and low SOC. Furthermore, we are also able to predict the topological character of materials without performing
any additional first-principles calculations, just by evaluating their position on the ``map''. By this means, we predict 74 additional novel QSHI candidates
in a distinct material class,~i.e.,~a set of different honeycomb lattice compounds with different functionalizations.


\section*{Results}

\subsection*{First-principles Classification of Functionalized 2D Honeycomb-Lattice Materials }
\begin{figure}[h!]
\center
\includegraphics[width = 0.55\linewidth]{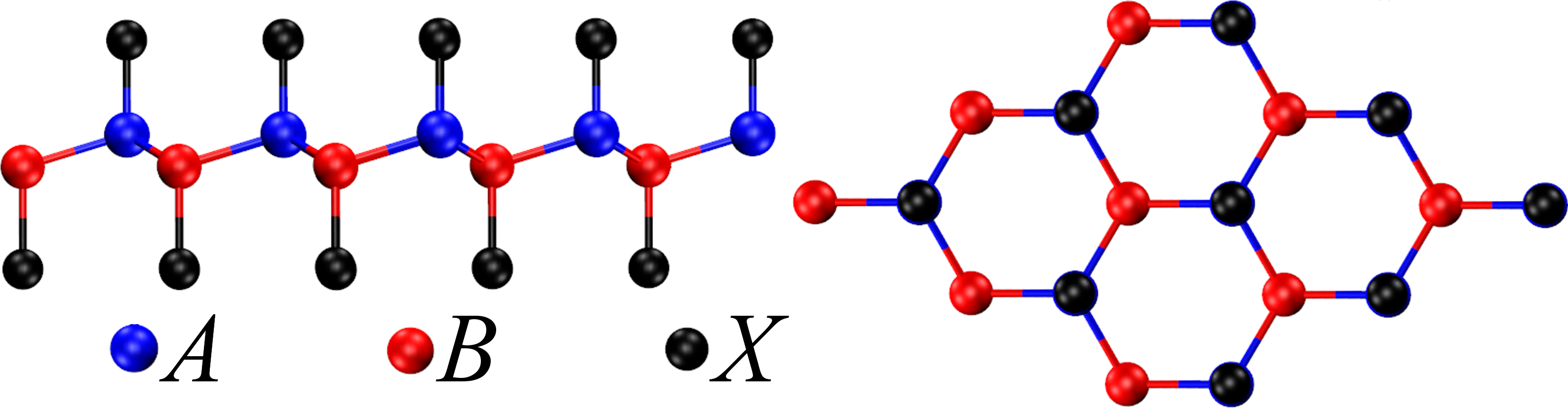}\\
\includegraphics[width = 0.55\linewidth]{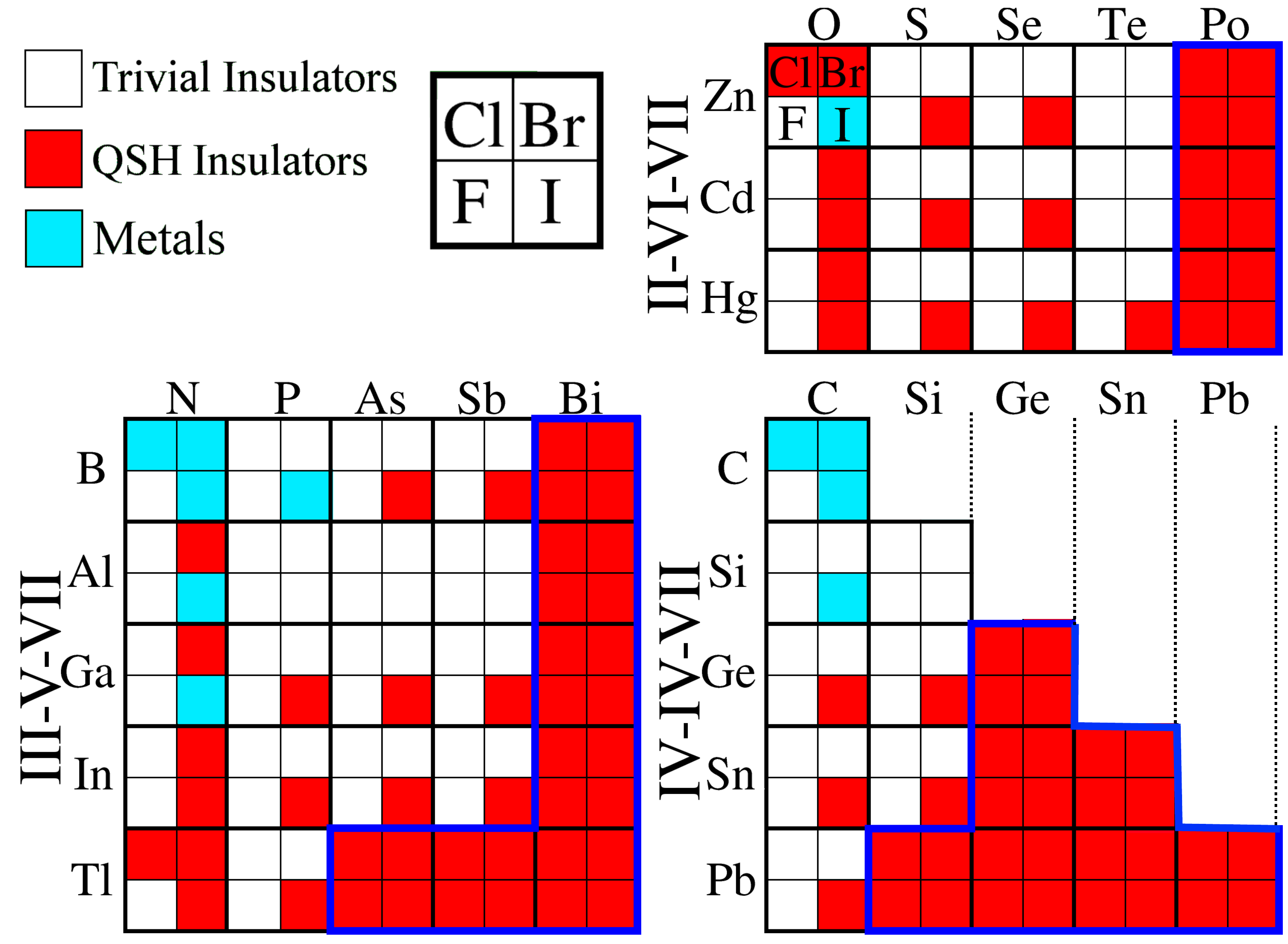}\\

\caption{Side and top view of the functionalized honeycomb-lattice system.
Classification~(trivial insulators: white; QSHIs: red; metals: cyan) of the 220 investigated $ABX_2$~compounds. The $x$ and $y$ axes denote the \textit{A} and \textit{B} atoms. For each combination~\textit{AB}, the four individual squares correspond to a different functionalization with a group VII element~(see legend). Compounds for which the topological character is independent of \textit{X} are surrounded by a blue line. See \SM for a tabulated list.}
\label{Geo}
\end{figure}

In a first step, we investigate the topological character of 220 functionalized 2D honeycomb-lattice materials~\textit{ABX}$_2$~(see Fig.~\ref{Geo})
by computing their $Z_2$ invariant from first principles. We include all possible combinations~\textit{AB} that are 
isoelectronic with graphene~(group IV-IV, III-V, and II-VI). For each honeycomb-lattice~\textit{AB}, functionalization with four different group VII~elements~(\textit{X} either Cl, Br, F, or I) 
is considered.
The resulting first-principles classification of these 220 compounds in metals~(zero band gap), trivial insulators~(nonzero band gap and $Z_2=0$), and QSHIs~(nonzero band gap and $Z_2=1$) 
is shown in Fig.~\ref{Geo}. 103 compounds are identified to be QSHIs in our calculations: In most cases~(66\%), the TI character is independent of the actual functionalization, as highlighted in blue in Fig.~\ref{Geo}. 
These 68 \underline{f}unctionalization-\underline{i}ndependent~(FI) QSHIs consist of relatively heavy elements, feature topological band gaps between 5~meV and 2~eV, and include 15 new QSHIs and 53 QSHIs reported in literature before,~e.g.,~functionalized stanene, germanene, Bi$_2$, GaBi, InBi, TlBi~\cite{nl500206u,PhysRevLett.111.136804,PhysRevB.89.115429,APL.1.4942380,PhysRevB.91.041303,PhysRevB.90.085431,
Song:2014hn,Jin:2015ie,nl504037u,nl504493d,Nano_Research.8.9.2954}. 
Additionally, we also identify 35 QSHIs~(34\% of all QSHIs), for which the TI character depends on the actual functionalization~(mostly iodides). These \underline{f}unctionalization-\underline{d}ependent~(FD) QSHIs 
with topological band gaps between 5~meV and 1~eV include 30 compounds that have not yet been reported in literature,~e.g.,~AlNBr$_{2}$ and GaAsI$_{2}$. 
Quite surprisingly, these TIs consist of relatively light elements with low SOC in the honeycomb lattice; they are however functionalized with relatively heavy atoms with strong SOC, which further substantiates that the topological transition
is driven by the functionalization. 

\subsection*{Descriptor Identification via Compressed Sensing}
To identify descriptors that can \emph{a priori} classify (i.e, the descriptor depends only on properties of the --- isolated --- atomic species constituting the material) functionalized 2D honeycomb-lattice materials in metals, trivial insulators, and QSHIs, 
we employed the SISSO~(\underline{s}ure \underline{i}ndependence \underline{s}creening and \underline{s}parsifying \underline{o}perator) approach 
recently developed by Ouyang {\em et al.}\cite{Runhai_paper}. First, a pool of about $10^{7}$ different potential descriptors~$D_{n}$ is constructed 
by analytically combining the properties of the free atoms \textit{A}, \textit{B} and \textit{X} computed with SOC~(namely, the eigenvalues of the highest-occupied and lowest-unoccupied 
Kohn-Sham states~$\Epsilon^{ho}$ and $\Epsilon^{lu}$, the atomic number $Z$, the electron affinity $\mbox{EA}$, the ionization potential $\mbox{IP}$, and the size $r_{s}$, $r_{p}$, and $r_{d}$
of the $s$, $p$, and $d$ orbitals, i.e., the radii where the radial probability density of the valence $s$, $p$, and $d$ orbitals are maximal. See \SM for a full list including the values.). 
Second, this compressed-sensing-based technique identifies which low-dimensional combination of these descriptors represents the classification best,~i.e.,~minimizes the overlap (or maximizes the separation)~\cite{acs.chemmater.5b04299}
among the convex hulls that envelope the individual classes~(metals, trivial insulators, QSHIs). This procedure, which is performed for a ``training'' set~(176 compounds 
randomly chosen from the total set of 220), reveals that the best descriptor for the classification of the investigated compounds is two-dimensional and features the components
\begin{eqnarray}
D_{1} & = & \left(Z_{A}+Z_{B}\right) \frac{\Epsilon^{ho}_{B}}{\mbox{EA}_{B}}\label{d1}\\
D_{2} & = & \mbox{EA}_{X}\,\mbox{IP}_{X}\,\left(r_{s,A}+r_{p,B}\right)\label{d2} \;.
\end{eqnarray}
For the remaining 20\% of compounds,~i.e.,~the so called ``test'' set used to validate the model,
we find that all materials with a very well defined structural and topological character are correctly classified. 
Only ZnOCl$_{2}$ and AlNBr$_{2}$, which are both FD-QSHIs with minute band gaps~($\le 15$~meV after SOC) 
at the verge of a topological transition to a trivial insulator or metal, respectively, are not correctly classified.
This shows that the found descriptor, which exhibits a predictive ability greater than 95\%, is robust and transferable,~i.e.,~not limited to the original training set used to identify~$D_{1}$ and $D_{2}$ 

\begin{figure}
\center
\includegraphics[width = 0.6\linewidth]{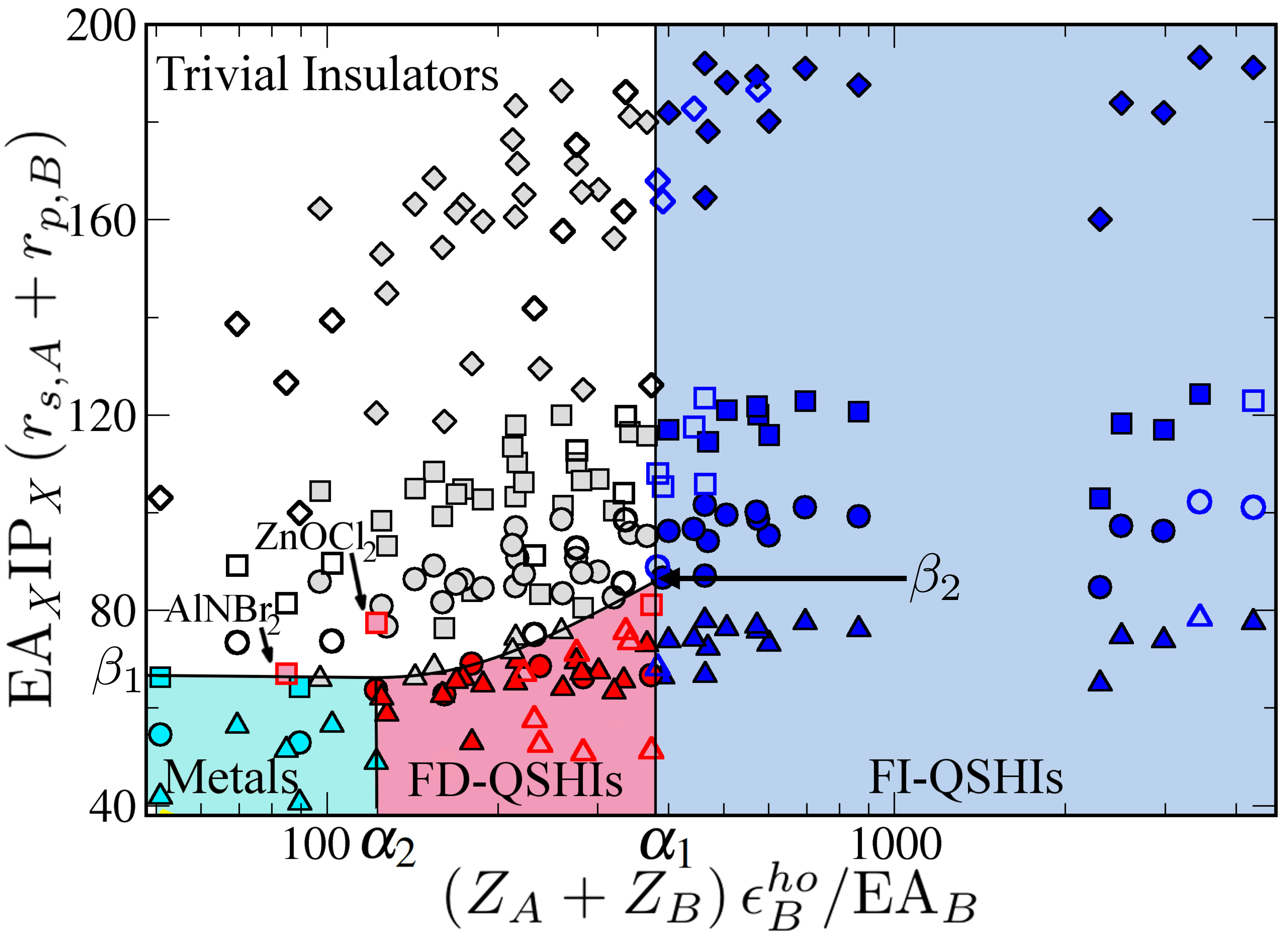}
\caption{Representation of DFT result for the training (filled) and test (unfilled symbols) set in the domain defined by the two-dimensional descriptors. A logarithmic scale is used for $D_{1}$. Compounds functionalized with F, Cl, Br, I are represented by diamonds, squares, circles and triangles, respectively. The symbols' color is used to distinguish between metalls~(cyan), FD-QSHIs~(red), FI-QSHIs~(blue), and trivial insulators~(white/grey). The same color-code is used to highlight the different regions identified by the SISSO descriptors. The boundaries of the map of materials are defined by $\alpha_{1}\approx 379$, $\alpha_{2}\approx 122.1$, $\beta_1\approx 70$, and $\beta_2\approx 85$.
The gap in the data points observed for $865< D_1 <2300$ is caused by the ``jump'' in $Z_A$ and $Z_B$ when switching from the 5$^\text{th}$ to the 6$^\text{th}$ row of the periodic system. See \SM for a tabulated list.}
\label{SISpred}
\end{figure}

As shown in Fig.~\ref{SISpred}, all compounds with $D_{1}>\alpha_{1}$ are FI-QSHIs~(blue). For~$D_{1}<\alpha_{1}$, the descriptor~$D_2$ matters as well: Trivial insulators~(white) occur for values of $D_2$ larger than the line connecting~$\beta_1$ and $\beta_2$, while for materials lying below that line we find metals~(cyan) in the region~$D_1<\alpha_{2}$ and FD-QSHIs~(red) for $\alpha_{2}<D_{1}<\alpha_{1}$. Note that $D_1$ and $D_2$ do not only clearly discriminate between metals, insulators, and QSHIs, but also separate FI-QSHIs from FD-QSHIs. 

\subsection*{Qualitative Interpretation of the Results}
\begin{figure}[h!]
\begin{center}
\includegraphics[width = 0.95\linewidth]{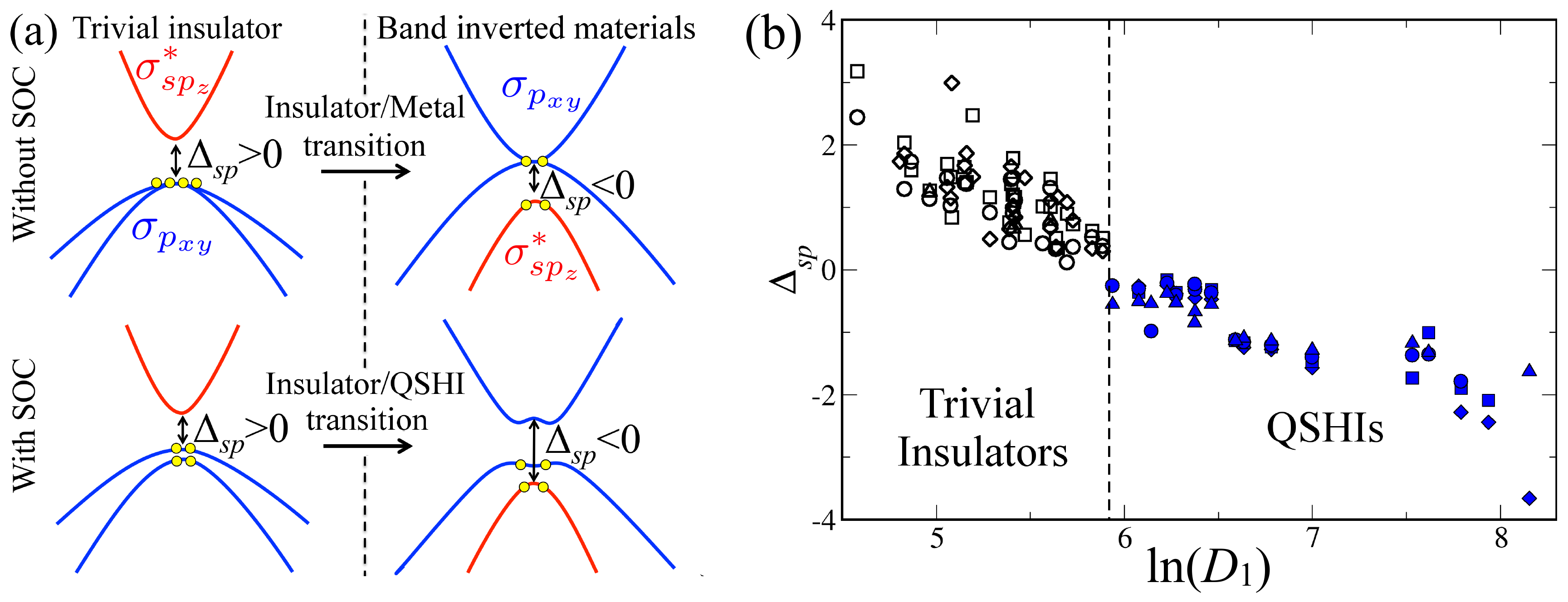}
\caption{(a) Representation of the band structure with and without SOC for trivial insulators and band-inverted materials.  Semimetals become insulators after including the SOC, leading to the QSHI phase. States formed by $p_{xy}$- (blue lines) and $sp_{z}$-orbitals (red lines) are inverted in the QSHIs; the energy distance between them ($\Delta_{sp}$) is related to the robustness of the topological state,~i.e.,~to which extent small perturbations such as strain can alter the toplogical character. The yellow dots represent four valence electrons; the four others accomodated in the bonding with the functionalization and the lower lying $\sigma_s$ state are not shown. (b) $\Delta_{sp}$ as a function of the logarithm of $D_{1}$ for all FI-QSHIs and trivial insulators.} 
\label{ProjectionVBM}
\end{center}
\end{figure}

The descriptor of components $D_{1}$ and $D_{2}$ does not only numerically and graphically sort the functionalized graphene-like materials, but also captures which atomic properties are relevant ``actuators'' for the topological phase transition. To understand this, it is necessary to clarify the character of the electronic 
states involved in the band inversion at $\Gamma$, as schematically sketched in Fig.~\ref{ProjectionVBM}a: These are the twofold degenerate $\sigma_{p_{xy}}$ state~(blue), to which 
the two in-plane valence $p_{xy}$-orbitals from atoms \textit{B} contribute, and the $\sigma_{sp_{z}}^*$ state~(red), to which the valence $s$ states from atoms~$A$ and the out-of-plane $p_{z}$ states from atoms $X$ contribute.  
Although all quantities entering Fig.~\ref{SISpred} are computed with SOC, the descriptor $(D_1, D_2)$ can be qualitatively rationalized as the relative energetic positions of these two states $\Delta_{sp}=E_{\sigma_{sp_{z}}^*}-E_{\sigma_{p_{xy}}}$ 
\textbf{before} SOC: If the twofold degenerate $\sigma_{p_{xy}}$ state lies lower in energy~($\Delta_{sp}>0$), it is fully occupied, since it accommodates all remaining  
valence electrons. In this case, we get a trivial insulator, since the SOC can only lift the degeneracy of the $\sigma_{p_{xy}}$~state, but not invert the band order. Conversely, the $\sigma_{sp_{z}}^*$ state lies lower 
in energy for $\Delta_{sp}<0$ and thus becomes fully occupied. With fewer remaining electrons, one thus gets a half-occupied $\sigma_{p_{xy}}$~state and so a band-inverted, semi-metal before SOC. The SOC itself is
thus solely responsible for the band-gap opening, which leads to the QSHI state. 

How the individual actuators present in the descriptor $(D_1, D_2)$ influence the topological transition, can be qualitatively understood from their influence on $\Delta_{sp}$.
As shown in Fig.~\ref{ProjectionVBM}b, $\Delta_{sp}$ correlates linearly with $\ln(D_1)$~($\Delta_{sp} \approx \left(-1.306\pm 0.048\right)\ln\left(D_1\right)+\left(8.015\pm 0.289\right)$)
for functionalization-independent QSHIs. In this case, the influence of the $p_z$~orbital of $X$ on $\sigma_{sp_{z}}^*$ is negligible, so that $\Delta_{sp}$ is dictated by atoms~$A$ and~$B$.
The role of the actuators in the component~$D_1$ of th descriptor can the be rationalized using the tight-binding models developed for tetrahedral group IV and III-V semiconductors~$AB$~\cite{AYQ:AYQA02705,PhysRevB.8.4487}. Here, ${\Epsilon^{ho}_{B}}/{\mbox{EA}_{B}}$ captures the energetic position of the hybridized $p_{xy}$-orbitals from atom \textit{B},
whereas~$\left(Z_{A}+Z_{B}\right)$ captures the size and thus the overlap of the respective orbitals. Just as in the case of tetrahedral semiconductors, heavier atoms lead to larger band widths
and thus to reduced band gaps~$\Delta_{sp}$ and metallic electronic structures.

Conversely, the component $D_2$ of the descriptor captures the influence of the functionalization. Due to its strong dependence on the electron affinity and ionization potential of atom~$X$, 
it groups the compounds by functionalization for $D_1>\alpha_{1}$, as apparent from Fig.~\ref{SISpred}. For $D_1<\alpha_{1}$, however, it describes the actual stable geometry 
of~$ABX_2$ and with that its electronic state: For light, strongly bound $AB$~compounds such as BN and~C$_2$~(small values of $Z_{A}+Z_{B}$ in $D_1$ and $r_{s,A}+r_{p,B}$ in $D_2$),
only strongly electronegative atoms $X$~(large values of $\mbox{EA}_{X}\,\mbox{IP}_{X}$) form a chemical bonding with $AB$, thus realizing a trivial insulators. Less electronegative 
functionalizations only physisorb via van-der-Waals interactions with $AB$, thus resulting in a metallic compound even after SOC. In this case, we also observe a structural transition 
  from the so called low-buckled~(LB) to the high-buckled structure~\cite{PhysRevB.90.241408}. The insulator/metal transition with $D_2<\beta_1$ and $D_1<\alpha_{2}$ is thus essentially a LB-HB structural phase transition that is energetically favorable only in these particular compounds, as we explicitly checked. In particular, this holds for compounds with~$D_1>\alpha_{2}$ 
that are relatively close to a topological transition from the start. Here, the additional degree of freedom provided by the functionalization allows to tune the $\sigma_{sp_{z}}^*$
state. For strongly electronegative atoms $X$~(large values of $\mbox{EA}_{X}\,\mbox{IP}_{X}$), this allows to close the gap $\Delta_{sp}$ and thus leads to a topological transition,
as discussed in the context of Fig.~\ref{ProjectionVBM}a before.

\begin{figure}
\center
\includegraphics[width = 0.6\linewidth]{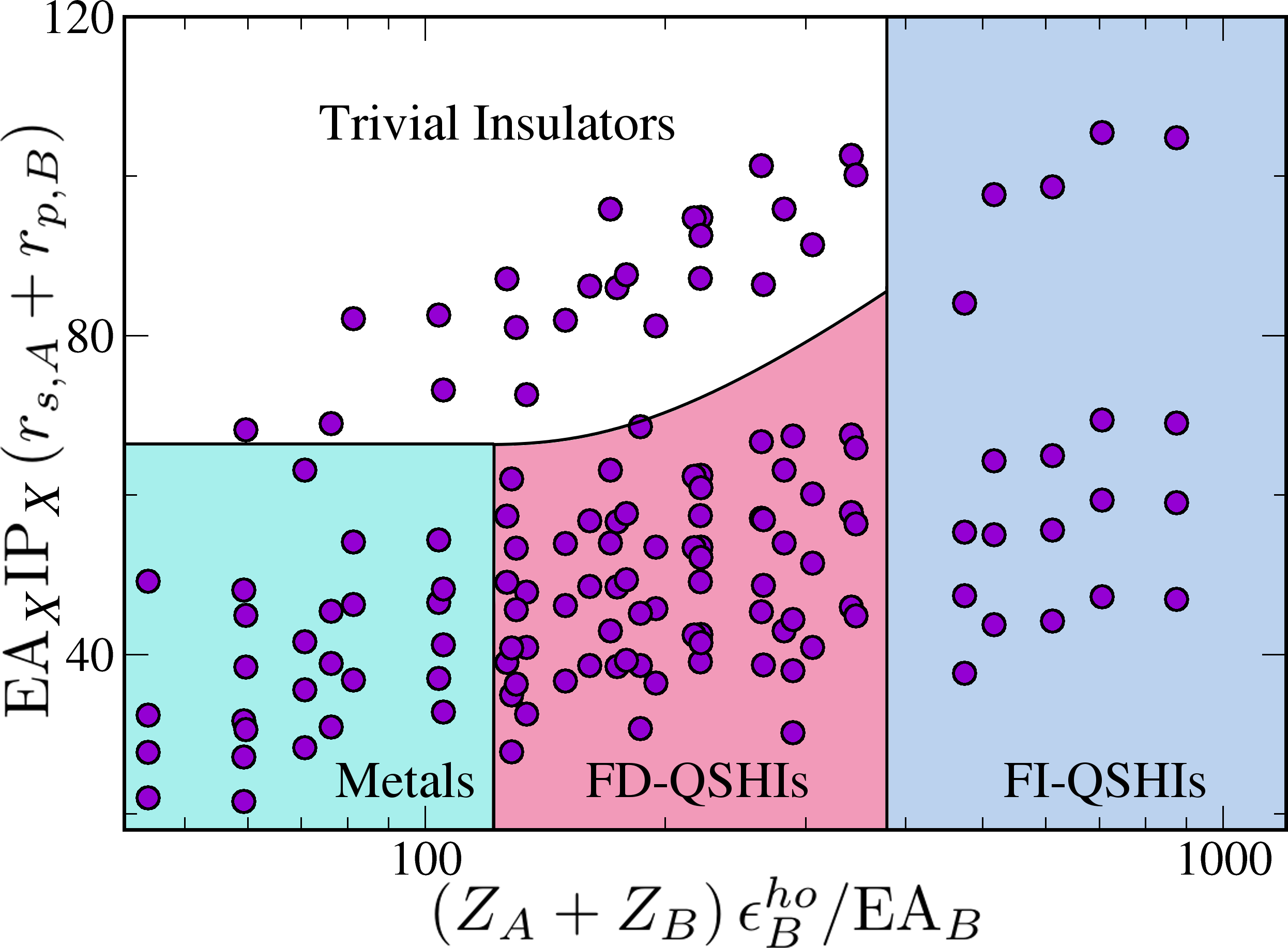}
\caption{
Representation of proposed V-V-VI and IV-VI-VI compounds in the domain defined by the two-dimensional descriptor components. 
See \SM for a full list of the materials and their classification.
}
\label{schematic}
\end{figure}

To showcase that the gained insights and identified descriptors are transferable, we have computed $D_{1}$ and $D_{2}$ for 140 less common honeycomb $ABX_2$ compounds~($AB$ from groups V-V and IV-VI functionalized with a  group VI element for $X$). For these compounds, the identified descriptor predicts 20 FI-QSHIs, 54 FD-QSHIs, 42 trivial insulators and 24 metals, as shown in Fig.~\ref{schematic} and tabulated in the Supp. Mat.. Since $\mbox{EA}_{x}\mbox{IP}_{X}$ for \textit{X} belonging to the VI group is lower, we get that the functionalization stabilizes the HB phase for non-QSHI materials. We have verified the prediction for selected compounds~(As$X$, Sb$X$, SnSe$X_2$, PSb$X_2$, Bi$X$ and PO with \textit{X}:O,S,Se,Te). 
It has been shown that oxide blue-phosphorene is a trivial insulator, which can become a QSHI by applying tensile strain~\cite{acs.jpcc.7b03808}. Recently, one compound has been reported as an intrinsic QSHI, AsO~\cite{Wang:2017el}.
Different from individual systems proposed by trial-and-error calculations, the employed SISSO approach allows to identify a complete family of 74 new QSHI candidates.

\section*{Methods}
\label{Methods}
\subsection*{First-principle calculations} For each of these systems, we have first determined the equilibrium lattice constant by relaxing both the atomic positions and the unit-cell shape until 
the residual forces on the atoms were smaller than 0.01 eV/\AA ~using the all electron, full potential numeric atom centered orbitals based electronic structure code {\it FHI-aims}~\cite{j.cpc.2009.06.022,j.jcp.2009.08.008,Knuth,science.aad3000}. 
For the equilibrium configuration, the topological invariant~$Z_2$ was computed from the evolution of the Wannier center of charge~\cite{Yu:2011fj,PhysRevB.83.235401,PhysRevB.83.035108,Gresch:2017hy} that we implemented in FHI-aims using the band structures and wavefunctions.  
For these latter properties, SOC was accounted for using a second-variational, second-order perturbation approach recently implemented in FHI-aims~\cite{Huhn:2017bp} and first used in Ref.~\cite{acs.chemmater.6b01832}. 
For a qualitative analysis of the band inversion mechanism, projected band structures were computed. 
All calculations were performed using the Perdew-Burke-Ernzenhof~(PBE) generalized gradient approximation\cite{PhysRevLett.77.3865}, the Tkatchenko-Scheffler van der Waals correction method~(DFT-TS)~\cite{PhysRevLett.102.073005}, and with numerical settings that guarantee a convergence of $<$1~meV for the eigenvalues. Specifically, periodic boundary conditions were used: the 2D hexagonal monolayers lie in the $xy$-plane, and a vacuum of 20\AA ~was used in the $z$-direction to avoid the undesirable interaction between the periodic images of sheets. Futhermore, ``really tight'' numerical settings and basis sets as well as a 40$\times$40$\times$1 $\vec{k}$ point grid for the Brillouin zone were used.

\subsection*{Statistical approach for identifying descriptors: Compressed Sensing}
To learn a descriptor for the $\mbox{Z}_2$-invariant material property, we employed the compressed-sensing approach recently developed by Ouyang {\em et al.}\cite{Runhai_paper}, which mainly consists of two steps: \textit{i}) construction of feature space (potential descriptors) by building analytical functions of the input parameters (atomic properties with SOC, in the case studied here), by iteratively applying a set of chosen algebraic operators, up to a certain complexity cutoff (number of applied operators). 
The used input atomic parameters are the eigenvalues of the highest occupied and lowest unoccupied Kohn-Sham states~$\Epsilon^{ho}$ $\Epsilon^{lu}$, the atomic number $Z$, the electron affinity $\mbox{EA}$, the ionization potential $\mbox{IP}$, and the size of the $s$, $p$, and $d$ orbitals ($r_{s}$, $r_{p}$, and $r_{d}$), i.e., the radii where the radial probability density of the valence $s$, $p$, and $d$ orbitals are maximal, for \textit{A}, \textit{B} and \textit{X}. 
All these features were computed using perturbative SOC~\cite{Huhn:2017bp}.
Consequently, the feature space is formed by $N$ vectors~$\boldsymbol{D}_{n}=(D_{n,1}, D_{n,2},...,D_{n,M})$, where $D_{n,m}$ is the $n^\text{th}$ combination of atomic features,~e.g.,~($\Epsilon^{ho}_{A}+\Epsilon^{ho}_{B}
+\Epsilon^{ho}_{X}$), evaluated on the constituent atoms of the $m^\text{th}$ \textit{ABX}$_{2}$ compound. For more details about the feature space construction please refer to Ref.~\cite{NewJPhys.19.023017,Runhai_paper}; \textit{ii}) descriptor identification by a scheme combining \underline{s}ure \underline{i}ndependence \underline{s}creening and \underline{s}parsifying \underline{o}perator, SISSO. 
SIS selects features $\boldsymbol{D}_{n}$, highly correlated with the $\mbox{Z}_2$ topological invariant property, which is formally written as a vector of the training values of $\mbox{Z}_{2}$-invariant.  
Starting from the features selected by SIS, the SO looks for the $\Omega$-tuples of features that minimizes the overlap (or maximize the separation)~\cite{acs.chemmater.5b04299},  
among convex hulls enveloping subsets of data. The dimensionality $\Omega$ of the representation is set as the minimal that yields perfect classification of all data in the ``training'' set. In this work $\Omega=2$ was found sufficient.  
This procedure is performed for a ``training'' set~(176 compounds randomly chosen from the total set of 220); the remaining 20\% are used as a ``test'' set to validate the found model.

\section*{Acknowledgments}
We thank the financial support by FAPESP under Grant Agreement No. 2016/04496-9 and the European Union’s Horizon 2020 research and innovation program under Grant Agreement No. 676580 with The Novel Materials Discovery (NOMAD) Laboratory, a European Center of Excellence. L.M.G. acknowledges funding from the Berlin Big-Data
Center (Grant Agreement No. 01IS14013E, BBDC). This project has received funding from the European Research Council (ERC) under the European Union’s Horizon 2020 research and innovation programme (Grant Agreement No. 740233, TEC1p).

\section*{Author contributions statement}
C.M.A. performed the first-principles calculations and SISSO classification, C.M.A. and C.C implemented the $\mbox{Z}_{2}$ invariant, R.O., M.S., and L.M.G developed the SISSO approach, R.O. implemented it, M.S. and A.F. conceived the idea, and C.C supervised the calculations. All authors have discussed the scientific results and have written the paper.

\bibliography{Ref} 
\end{document}